\documentclass[11pt]{article}

\usepackage[letterpaper, portrait, margin=0.8in]{geometry}
\usepackage{amssymb,amsmath}

\usepackage{graphicx} 
\usepackage[margin=0.3in]{caption}
\usepackage{subfigure}
\usepackage{soul}
\usepackage{mdframed}
\usepackage{booktabs}

\usepackage{sectsty}
\usepackage{mathrsfs}
\usepackage{bm}
\usepackage[sort&compress,numbers]{natbib}

\DeclareMathAlphabet\mathbfcal{OMS}{cmsy}{b}{n}
 
\title{Cluster Amplitudes and Their Interplay with Self-Consistency in Density Functional Methods}

\author{Greta Jacobson,${}^{1,2}$ Juan M. Marmolejo-Tejada,${}^1$ Mart\'in A. Mosquera${}^{1,*}$\\
\small ${}^1$Department of Chemistry and Biochemistry, Montana State University, Bozeman, Montana 59717 USA\\
\small ${}^2$Department of Chemistry, Millikin University, Decatur, Illinois 62522 USA\\
\small ${}^*$ \texttt{martinmosquera@montana.edu}
}

\date{August, 2022}

\begin{document}
\newtheorem{thm}{Theorem}
\newtheorem{cor}{Corollary}

\def\bea{\begin{eqnarray}}
\def\eea{\end{eqnarray}}
\def\ben{\begin{equation}}
\def\een{\end{equation}}
\def\benu{\begin{enumerate}}
\def\enu{\end{enumerate}}

\newcommand{\mr}[1]{\mathrm{#1}}
\newcommand{\mc}[1]{\mathcal{#1}}
\newcommand{\mb}[1]{\mathbf{#1}}
\newcommand{\lket}[1]{\langle #1|}
\newcommand{\rket}[1]{| #1\rangle}
\newcommand{\ud}{\mr{d}}
\newcommand{\ui}{\mr{i}}
\newcommand{\intdr}{\int \ud^3\mb{r}~}
\newcommand{\XC}{\mr{XC}}
\newcommand{\sA}{_\mr{A}}
\newcommand{\sB}{_\mr{B}}
\newcommand{\XX}{\mr{X}}
\newcommand{\cc}{\mr{c}}
\newcommand{\inL}{i_\mr{L}}
\newcommand{\jnL}{j_\mr{L}}
\newcommand{\HXC}{\mr{HXC}}
\newcommand{\GS}{\mr{GS}}
\newcommand{\LDA}{\mr{LDA}}
\newcommand{\GGA}{\mr{GGA}}
\newcommand{\LSDA}{\mr{LSDA}}
\newcommand{\ps}{\mr{ps}}
\newcommand{\h}{_{\mr{h}}}
\newcommand{\Ha}{\mr{H}}
\newcommand{\upa}{\uparrow}
\newcommand{\dwna}{\downarrow}
\newcommand{\s}{\mr{s}}
\newcommand{\dernr}[1]{\frac{\delta {#1} }{\delta n(\mb{r})}}
\newcommand{\dernrs}[1]{\frac{\delta {#1}}{\delta n_{\sigma}(\mb{x})}}
\newcommand{\derN}[1]{\frac{\partial {#1}}{\partial N}}
\newcommand{\eel}{$e^{-}$}


\maketitle

\abstract{
Density functional theory (DFT) provides convenient electronic structure methods for the study of
molecular systems and materials. Regular Kohn-Sham DFT calculations rely on unitary transformations
to determine the ground-state electronic density, ground state energy, and related properties.
However, for dissociation of molecular systems into open-shell fragments, due to the
self-interaction error present in a large number of density functional approximations, the
self-consistent procedure based on the this type of transformation gives rise to the well-known
charge delocalization problem. To avoid this issue, we showed previously that the cluster operator
of coupled-cluster theory can be utilized within the context of DFT to solve in an alternative and
approximate fashion the ground-state self-consistent problem. This work further examines the
application of the singles cluster operator to molecular ground state calculations. Two
approximations are derived and explored: i), A linearized scheme of the quadratic equation used to
determine the cluster amplitudes, and, ii), the effect of carrying the calculations in a
non-self-consistent field fashion. These approaches are found to be capable of improving the energy
and density of the system and are quite stable in either case. The theoretical framework discussed
in this work could be used to describe, with an added flexibility, quantum systems that display
challenging features and require expanded theoretical methods.
}

\section{Introduction}

Electronic structure methods predict a very large number of measurable quantities that are used to
understand, characterize, and optimize chemical compounds and materials. Quantum mechanics is the
foundation upon which algorithms are designed and applied to compute electronic and structural
properties. From a fundamental standpoint, quantum mechanics states that with a complete knowledge
of the wave function of the system one can thus be able to determine all the information about the
system of interest. For computational efficiency, however, density functional theory (DFT) serves as
an alternative to pursue such goal. In DFT one the primary objectives is the calculation of the
electronic density of the system, as opposed to the full wave function of all the electrons.
Although it is common to separate both, wave-function theory (WFT) and DFT, as separate fields, it
can be argued that both are intrinsically connected, especially from the algorithmic point of view. 

DFT methods have been formulated on the basis of physical understanding of model systems and small
molecules. A notable example is the electron gas, which in many ways has led to functional
components that to date still remain an important part of a very large number of density functional
approximations (DFAs). These functionals are available for different energy ``pieces'' such as the
kinetic, exchange, correlation, and van der Waals energies. The kinetic energy is known to be the
most challenging energy to be expressed explicitly as a density-functional. For this reason,
Kohn-Sham (KS) DFT \cite{kohn1965self} is the most common theory within DFT that is utilized for
practical calculations and to derive concepts.\cite{burke2012perspective,chamorro2021chemical} As
KS-DFT uses single-electron orbitals to determine a kinetic energy. As is well known, even though
KS-DFT practical calculations perform well for determining properties such as molecular geometries,
and optoelectronic properties of a very large number of compound types, it is difficult for
transition-metal systems \cite{siegbahn2010significant}, bond-breaking \cite{perdew1982density}, and
charge-transfer excitations \cite{mester2022charge}, among others, where erroneous charge
delocalization
\cite{matito2009role,lundberg2005quantifying,otero2014halogen,autschbach2014delocalization} is a
main manifestation of these adverse effects.

Extended DFAs that are free of incorrect charge delocalization should eliminate the main cause for
such adverse effect, the self-interaction error
\cite{zhang1998challenge,mori2006many,vydrov2007tests,cohen2008insights}. Additionally, improved
methodologies must also come with relatively low computational costs.  Motivated by these
considerations, and fueled by advances in machine learning and the premise of new generation of
computing technologies (classical and quantum), theoretical methods are being advanced by the
scientific community, with the goal of extending the applicability of DFT methods
\cite{dick2020machine,pederson2022machine,kirkpatrick2021pushing}. These extensions include the
development of force fields, which are creating opportunities for detailed studies of systems at the
mesoscopic scale \cite{ nochebuena2021development}. For example, artificial neural network (ANN)
algorithms have been used to generate density functional approximations \cite{Custodio2019,Zhu2019},
and have been able to eliminate charge delocalization errors. On the other hand, ANNs also have led
to both transferable and specific force fields. This also includes ANNs being used extensively in
materials discovery and properties prediction \cite{Gubaev2019,Nyshadham2019,Zuo2020}.
Machine-learned interactomic potentials, which are tailored for a particular system of interest
demonstrate quite appealing theoretical prospects for modeling mesoscale phenomena
\cite{Shapeev2016,Behler2016,Novoselov2019,Ladygin2020,Mortazavi2020,Mortazavi2020a,Marmolejo-Tejada2022}. 

From a foundational perspective, the elimination of charge delocalization still remains a long
sought goal, where theoretical tools are still the subject of continued developments. This problem
not only manifests in DFT development, but also in WFT research. For example, it is known that there
are dynamically correlated post-Hartree-Fock methods that can also cause issues with
size-consistency, whereas the well-known exponential ansatz of WFT, in conjunction with
spin-symmetry breaking, offers a theoretically sound route to restore size-consistency (which
implies size-extensivity as well). We showed previously that this exponential operator, which in
turn is determined by what is known as the ``cluster operator''
\cite{coester1958bound,coester1960short,vcivzek1966correlation,vcivzek1969use,monkhorst1977calculation,mukherjee1979response,emrich1981extension,
emrich1981extension2,ghosh1984use,stanton1993equation,bartlett2007coupled}, can also prevent
undesired charge delocalization in DFT calculations \cite{mosquera2021density}. The cluster operator
in the ground-state case is limited in our calculations to single-electron transitions, as it
displays a high degree of accuracy at this level of excitation. The cluster amplitudes that are used
to construct the exponential operator are derived as the solution of a quadratic equation, which is
solved in an iterative fashion. Our proposed method, denoted as ``eXp'' (due to its relying on the
exponential operator), predicted with physical consistency the binding energy curves of classical
systems such as di-hydrogen, lithium hydride, and hydrogen fluoride, but we also show other cases
where the eXp method functions as an alternative to the standard unitary method of KS-DFAs, and we
suggested they are also compatible with the double-hybrid functional approach
\cite{goerigk2014double,chai2009long,schwabe2007double}. These previous findings motivate the
present work, where we further explore the eXp method under its linearized version, which simplifies
in a very accurate way the determination of the cluster amplitudes and the exponential operator. We
also examine non-self-consistent field calculations, where the single-particle Hamiltonian is
determined by the Hartree-Fock density, which is used to estimate directly the cluster operator and
its conjugate, the ``lambda'' operator. In this study we find that the linearized eXp method
performs quite well with excellent agreement with respect to the full quadratic scheme in both
cases, the self-consistent and the non-self-consistent ones. The eXp technique is applied to a
couple of known cases of severe charge delocalization (or strong self-interaction), with the goal of
eliminating it: The positively charged neon dimer, Ne$_2^+$, and lithium-fluoride, LiF. In addition,
our methods are applied to a set of molecules at their minimum-energy geometries, where we show that
the linearized eXp method performs quite similarly as the quadratic version in self-consistent-field
(SCF) and {\slshape non}-self-consistent-field (NSCF) calculations. However, the NSCF computations, as expected,
are less accurate that the SCF ones, but can be considered for calculations where computational
acceleration is needed. The simulations considered in this work are based on a single-particle
Hamiltonian, but they are also applicable to Hamiltonians that include two-body interactions, such
as those used in double-hybrid approaches.


\section{Theory}

\section{Computational Details}

Determining ground-state properties in KS-DFT begins with the calculation of the KS Slater
determinant $|\Phi\rangle$ and subsequently the electronic energy. The wave function $|\Phi\rangle$
is computed through the minimization of an auxiliary single-particle energy, which depends on the
single-particle Hamilonian, or KS Fock operator. We denote this density-dependent operator as
$\hat{f}$. The energy function that is minimized in KS-DFT to obtain the orbitals is then
$\langle\Phi|\hat{f}|\Phi\rangle$, and it leads to the standard KS equations where the single
particle orbitals are constructed through diagonalization of the KS Fock matrix. The object
$\hat{f}$ is the sum of the kinetic, electron-nucleus, exchange-correlation (XC), and Hartree
contributions.

As an alternative to the standard procedure mentioned above, we stationarize the single-particle
energy with respect to cluster operators, where the reference is a Hartree-Fock (HF) wave function,
which we denote as $\Psi_0$. This wavefunction, as expected, is constructed with occupied orbitals
in the HF molecular orbital basis set. This is a relevant detail, as our calculations rely entirely
on such molecular basis set.  The HF wavefunction can either be a restricted or unrestricted
reference. We introduce an auxiliary right-handed wave function of the form
$|\Upsilon_{\mr{R}}\rangle=\exp(+\hat{t})|\Psi_0\rangle$, and the left-ket $\langle
\Upsilon_{\mr{L}}|=\langle\Psi_0|(1+\hat{\Lambda})\exp(-\hat{t})$, where $\hat{t}$ and
$\hat{\Lambda}$ are the cluster operators. The function to stationarize is
$\langle\Upsilon_{\mr{L}}|\hat{f}|\Upsilon_{\mr{R}}\rangle$, so it leads to the auxiliary
single-particle energy as:
\begin{equation}
E_{\mr{s}}=\underset{\hat{t},\hat{\Lambda}}{\mr{stat.}}\langle\Psi_0|(1+\hat{\Lambda})\bar{f}|\Psi_0\rangle
\end{equation}
\sloppy where the symbol $\bar{f}$ denotes the transformed operator $\exp(-\hat{t})\hat{f}\exp(+\hat{t})$.
We use this notation for other operators too; so if $\hat{\Omega}$ is some arbitrary operator, then
$\bar{\Omega}=\exp(-\hat{t})\hat{\Omega}\exp(+\hat{t})$. The cluster operators that we are
interested in have the form $\hat{t}=\sum_{ai}t_i^a \hat{a}^{\dagger}\hat{i}$, and
$\hat{\Lambda}=\sum_{ai}\Lambda_i^a\hat{i}^{\dagger}\hat{a}$. The indices $i$ and $a$ denote
occupied and virtual spin-orbitals, respectively. By stationarizing with respect to $\hat{t}$ and
$\hat{\Lambda}$ it is then implied that one must find, what we regard as vectors computationally,
$\{t_i^a\}$ and $\{\Lambda_i^a\}$. This demands that the derivatives of the function
$\langle\Psi_0|(1+\hat{\Lambda})\bar{f}|\Psi_0\rangle$ with respect to all the elements
$\Lambda^a_i$ and $t^a_i$ are all zero.

We denote $f_{pq}$ as the matrix element, $\langle \chi_p|\hat{f}|\chi_q\rangle$, where $\chi_p$ is
a {\slshape Hartree-Fock spin-orbital}; this implies that $f_{pq}$ can be non-zero for $p\neq q$. We
then have that the $t$-amplitudes derive from the equation:
\begin{equation}\label{maineq}
0=f_{ai}+\sum_b t^b_i f_{ab}-\sum_j t_j^af_{ji}-\sum_{jb}f_{jb}t_i^bt_j^a
\end{equation}
And the $\Lambda$-amplitudes are obtained from the linear system $\mb{M}\bm{\Lambda}=-\bm{f}$, where
\begin{equation}
    M_{ck,ai}=R_{ck,ai}-\sum_j t_j^af_{jc}-\sum_bt_i^b f_{kb}\delta_{ac}
\end{equation}
and
\begin{equation}
    R_{ck,ai}=f_{ca}\delta_{ik}-f_{ki}\delta_{ca}
\end{equation}
The symbol $\mb{f}$ represents the Fock matrix as a vector, $(\mb{f})_{ai}=f_{ai}$. We denote the
process of determining $\mb{t}$ through Eq. \ref{maineq} as the quadratic eXp scheme, or ``Q-eXp''.
It, Q-eXp, can be solved using the quasi-Newton method where an estimate to $t_i^a$ is updated
according to the equation:
\begin{equation}\label{t_iter}
    t_i^a\leftarrow t_i^a - \frac{L_i^a}{f_{aa}-f_{ii}}
\end{equation}
Where $L_i^a$ refers to the left-hand side of Equation \ref{maineq}.

By neglecting quadratic terms in Eq. \ref{maineq}, we obtain the approximation:
\begin{equation}\label{lexp}
    \mb{R}\mb{t}=-\mb{f}
\end{equation}
We refer to this scheme as ``L-eXp''. This approximation requires the solution to a linear system of
equations, so it avoids the need for iterations to find $t$. On the other hand, this linear matrix
equation can be further reduced to the simple, approximated, analytical expression: $t^a_i =
-f_{ai}/(f_{aa}-f_{ii})$, which we used before as a first estimate to start a the iterative cycle in
Q-eXp.  In this work we explore the L-eXp and Q-eXp methods in NSCF and SCF procedures. So NSCF
L-eXp, for instance, refers to the use of the linearized eXp method, Eq. \ref{lexp}, where the
amplitudes are computed only one time, and the XC and Hartree potentials are evaluated at the
Hartree-Fock densities; the same applies to NSCF Q-eXp.

\begin{figure}[thb]
\begin{center}
\includegraphics[scale=0.65]{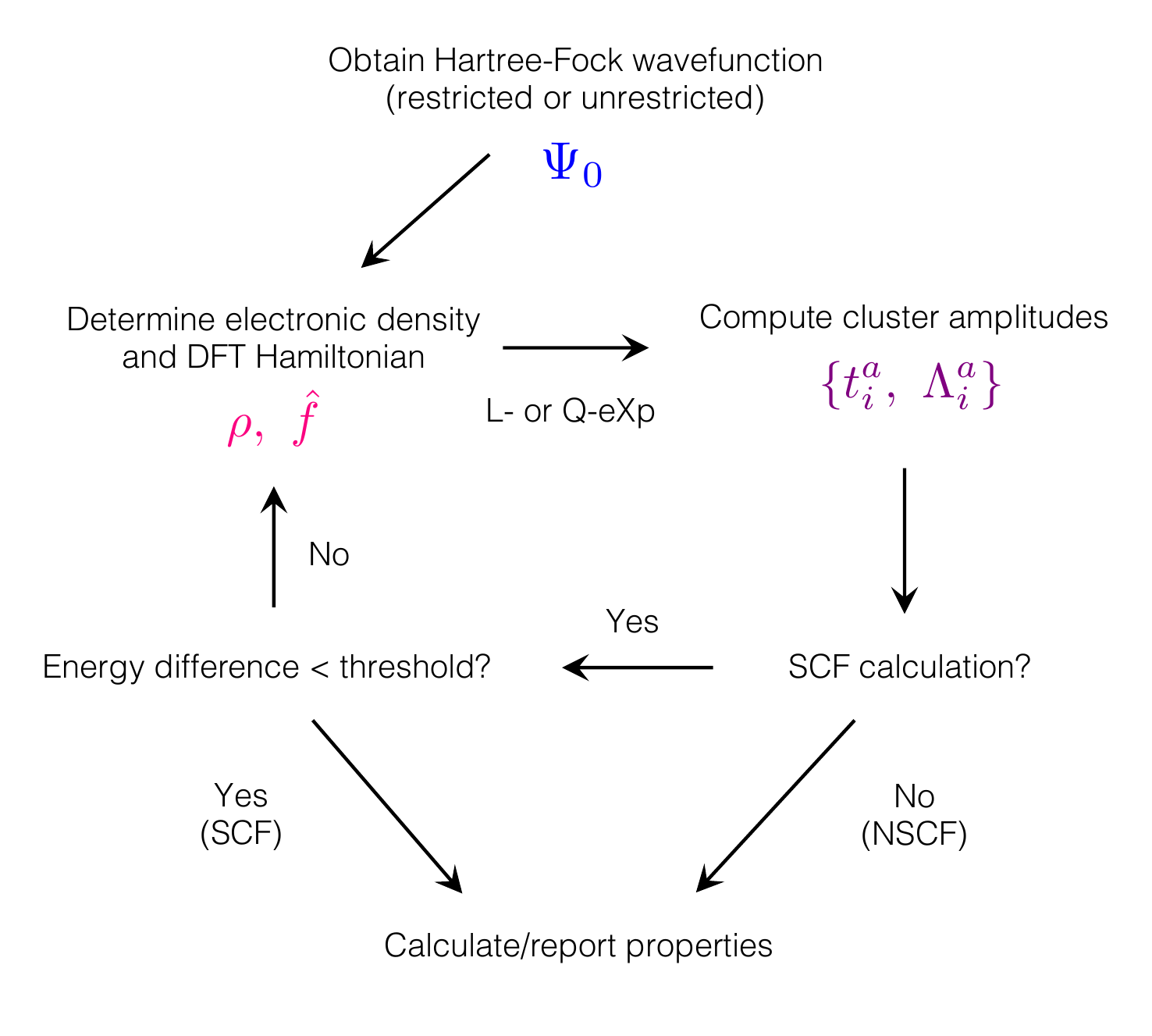}
\end{center}
\caption{Flow chart summarizing steps carried out in an single-point calculation based on an eXp
method (linear or quadratic). The cluster amplitudes are determined with respect to the Hartree-Fock
reference and molecular basis.}
\label{flowc}
\end{figure}

Given the solution to the above problems the energy is calculated as:
\begin{equation}
E=\langle\Psi_0|(1+\hat{\Lambda})e^{-\hat{t}}\hat{h}_0e^{+\hat{t}}|\Psi_0\rangle+E_{\mr{Hxc}}[\rho]    
\end{equation}
where $\hat{h}_0$ is the core Hamiltonian (kinetic plus electron-nuclei attraction energy
operators), and $\rho(\mb{r})=\langle (1+\hat{\Lambda})\bar{\rho}(\mb{r})\rangle$. The term
$E_{\mr{Hxc}}$ refers to the sum of the Hartree and XC energies, where the XC energy is approximated
with a DFA. The steps followed to calculate the ground-state energy, and related properties are
summarized in Figure \ref{flowc}. As usual, in the SCF cycle the electronic density is updated until
the energy variation between iterations is below a certain threshold. In the NSCF approach the
cluster amplitudes are only determined one time, with the Fock operator being based on the (U)HF
electronic density, or density matrix if the XC functional is hybridized. 

An important quantity in our calculation is the fundamental energy gap of the system, as our iterations
depend on differences of the type $f_{aa}-f_{ii}$, at moderately long distances a few of these can be close
to zero, which cause instabilities. To eliminate them, we use a regularization scheme in which the
Fock operator is modified by the $t$ cluster amplitudes, so the new operator is
$\hat{f}_{\alpha}=\hat{f}+\alpha\hat{t}$, where the regularization number $\alpha > 0$, and the
problem is solved with respect to such single-particle Hamiltonian, otherwise the methodology
remains the same. In Ref. \cite{mosquera2021density} we show details of this regularization
procedure. For the quadratic method Q-eXp, in Eq. (\ref{t_iter}) the difference $f_{aa}-f_{ii}$ is
replaced by $f_{aa}-f_{ii}+\alpha$ and the term $L_i^a$ by $L_i^a+\alpha t_i^a$. For the linearized
eXp method, we just add the constant $\alpha$ to all the diagonal elements of the matrix $\mb{R}$.
Around minimum-energy, or equilibrium, geometries we do not find a need to use such regularization
scheme, but there are other cases where this is necessary. Regularization is a benign procedure that
eliminates instabilities and is used in standard coupled-cluster
\cite{kowalski2009extensive,kowalski2009generating,taube2009rethinking,lawler2008penalty},
perturbative theories \cite{song2017analytical,bertels2019third}, multireference methods
\cite{battaglia2022regularized}, and related theories such as pseudo-potentials and
machine-learning.

\section{Computational Details}
The calculations presented in this work were run using a series of python scripts based on the
PyQuante suite \cite{pyq}. The local spin-density approximation (LSDA) is used in pure and
hybridized forms. Two hybrids of interest are considered, the ``half-and-half'' one, consisting of
50 \% HF exchange, 50 \% LSDA exchange, and 100 \% LSDA correlation energies, we refer to this
functional as LSDA-H. The second hybridized functional is denoted as ``LSDA-75'', consisting of 75 \% HF
exchange, 25 \% LSDA exchange, and 100 \% correlation energies. All our bond-dissociation
calculations were performed with the 6-31++G** basis set. The convergence threshold for the
(unrestricted) HF calculations is $10^{-8}$ au, and for the $t$-amplitudes in the Q-eXp case
$10^{-6}$. Tighter thresholds are possible, but were not needed in our simulations.  The SG-2 grid
is used to represent the XC potential and energy-density and to compute the XC energy. Reference
calculations were performed with the Q-Chem computational chemistry software\cite{shao2015advances}
for the standard KS calculations with the LSDA-H and -75 functionals, which are built using its
user-defined density-functional interface. Reference unrestricted coupled cluster singles and
doubles (UCCSD) were also carried out with Q-Chem.  For Table \ref{tab:enerdipoles} shown in
next section, the minimum-energy geometries derive from MP2 calculations using aug-cc-pVQZ
calculations that were performed with the NWChem program \cite{nwchem}.

\section{Results and Discussion}

\begin{figure}[thb]
\begin{center}
\subfigure[$\mr{Ne}_2^+$ binding energy curve]{\includegraphics[scale=0.4]{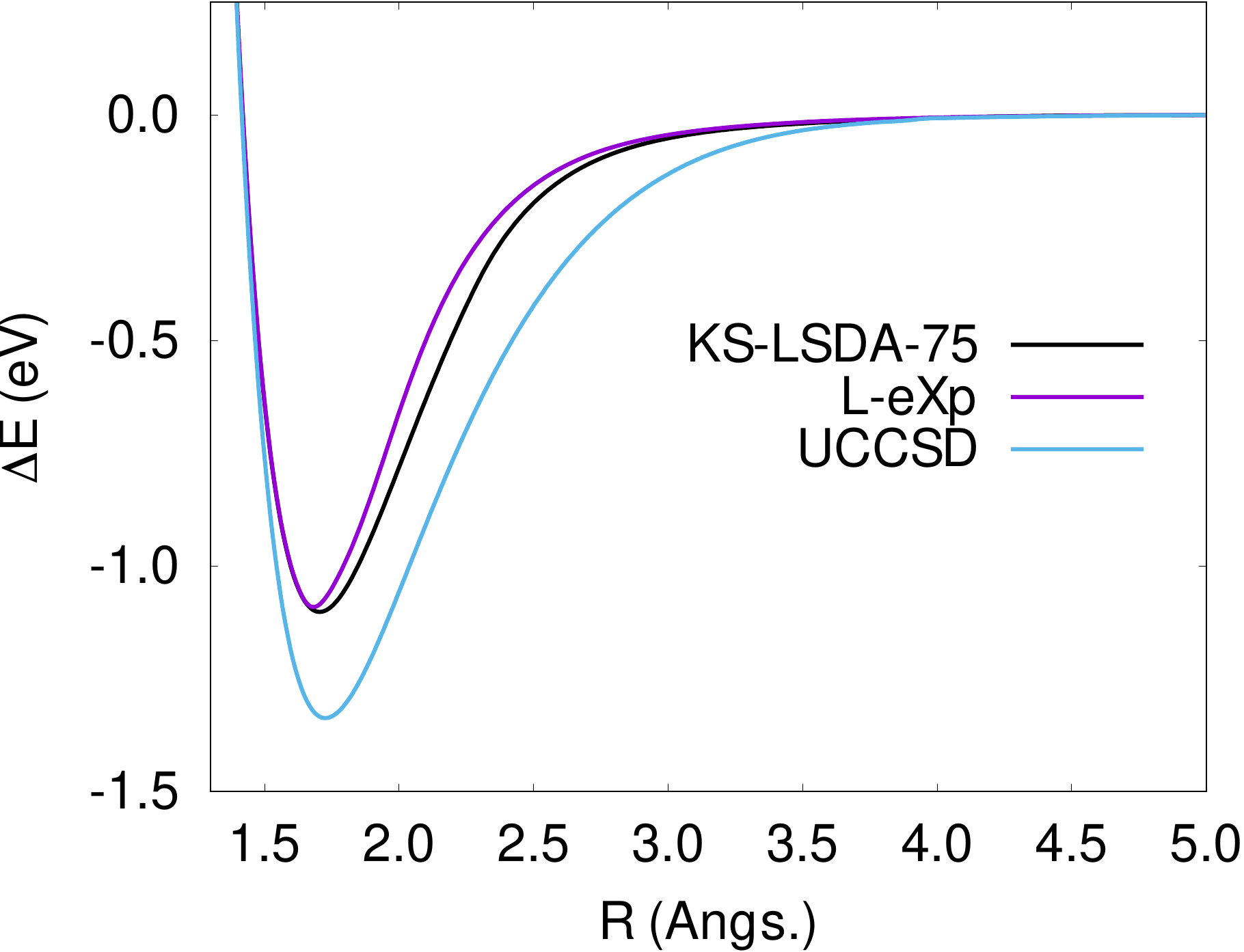}}
\subfigure[NSCF vs SCF methods]{\includegraphics[scale=0.4]{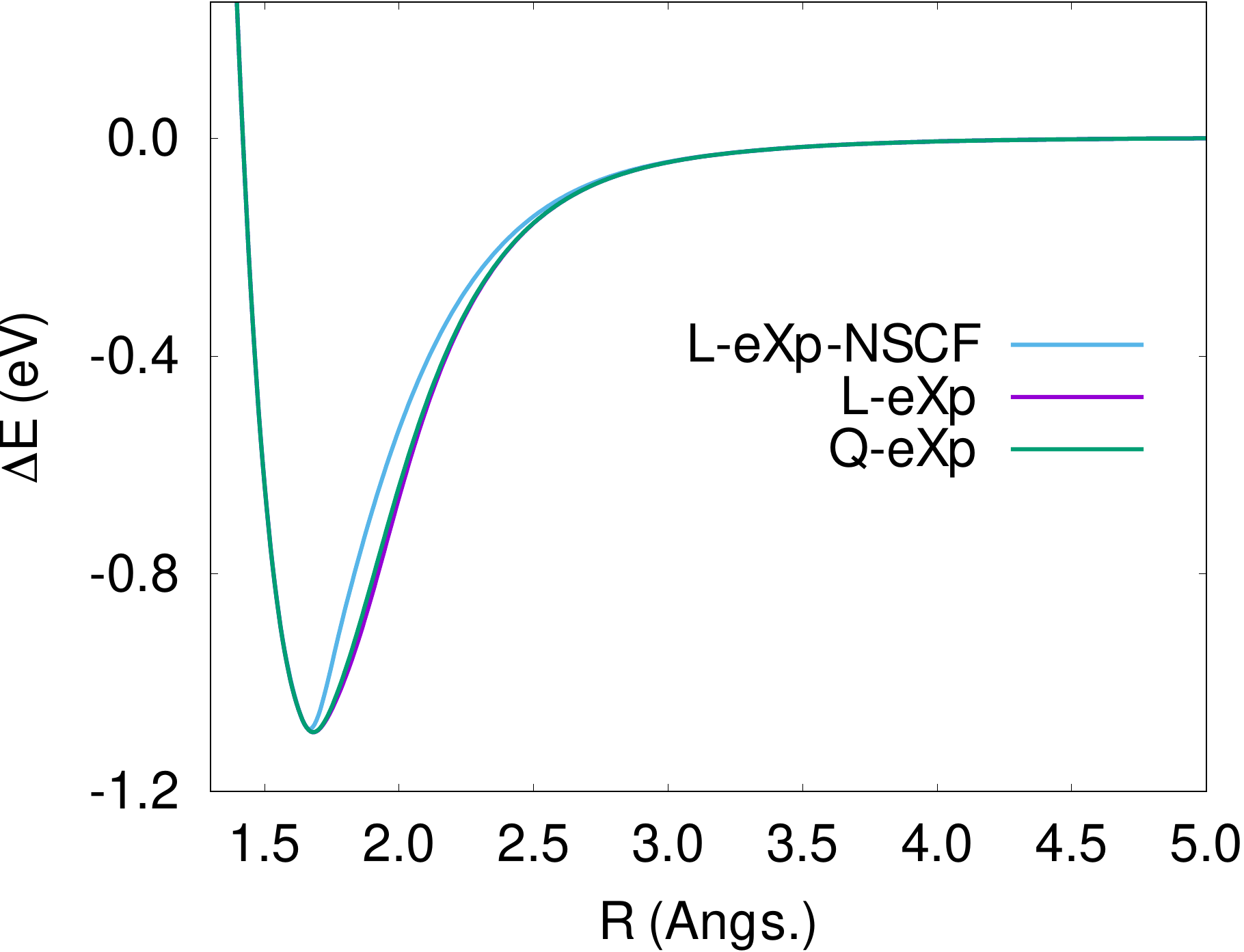}}
\end{center}
\caption{Performance of standard and eXp methods for the binding energy curve of Ne$_2^+$, where $R$
denotes the internuclear distance: a), Comparison between a standard hybridized KS-LSDA calculation,
self-consistent linearized eXp, and UCCSD. This curve shows that the linearized eXp calculation
closely follows that of the standard KS-LSDA hybrid. b), Comparison between linearized and quadratic
self-consistent eXp computations and a linearized non-self consistent simulation (all with
$\alpha=0.1$ au). The NSCF calculation works well at equilibrium, repulsion, and dissociation, with
some relatively small differences otherwise.}
\label{scf_comparison_ne2p}
\end{figure}

We begin our discussion with the Ne$_2^+$ system. At dissociation of this diatomic molecule, the
unrestricted Hartree-Fock (UHF) spin-symmetry process yields one atom as being neutral and the other
one with positive charge. However, with a density functional such as the purely density-dependent XC
LSDA, the energy levels of the atoms display an undesired behavior from the point of view of spin
symmetry breaking: The lowest unoccupied $p$ spin-level of the cation lies below that of the
occupied $p$-shell of the neutral atom by about 18 eV, so the SCF algorithm will bias the
ground-state minimization toward a charge-transfer configuration (where an electron is shared
between the two atoms), delocalizing the positive charge and unphysically lowering the energy, as
the $\mr{Ne}-\mr{Ne}^+$ (neutral-cation) configuration energy configuration has a higher energy. The
eXp method we propose can eliminate this problem in a pure LSDA calculation, but it requires
strong regularization for moderately long distances between atoms. As mentioned before, which is well-known
in the literature, the cause for charge delocalization is the self-interaction error. Therefore, it
is possible to add HF exchange until the standard KS method performs appropriately at dissociation.
This happens, for example, when the amount of HF exchange is 75 \%, as shown in Figure
\ref{scf_comparison_ne2p}.a, where our linearized eXp method reaches a size-consistent result, as
well as the standard spin-symmetry-broken KS-LSDA-75 method. At this hybridization strength,
however, the binding energy is underestimated with respect to UCCSD, which is more reliable in this
case. But the L-eXp SCF result follows closely the standard KS-LSDA-75. Even at this level of
hybridization, nevertheless, there are differences in energy level that are close to zero, when this
occurs there are instabilities in the cluster amplitudes. For this reason, our calculation, L-eXp,
includes a regularization number of 0.1 au. In previous work \cite{mosquera2021density} we showed
these eliminate iterative divergences while maintaining physical consistency with the parent methods
used for comparison.  As highlighted previously, an eXp calculation can proceed in a self-consistent
fashion or not. In Figure \ref{scf_comparison_ne2p}.b we show that the linear and quadratic eXp SCF
approaches yield very similar results, whereas the linearized eXp method shows some deviations, but
it remains physically meaningful with respect to the SCF calculations.

\begin{figure}[thb]
\begin{center}
\subfigure[$\mr{Ne}_2^+$ binding energy curve]{\includegraphics[scale=0.4]{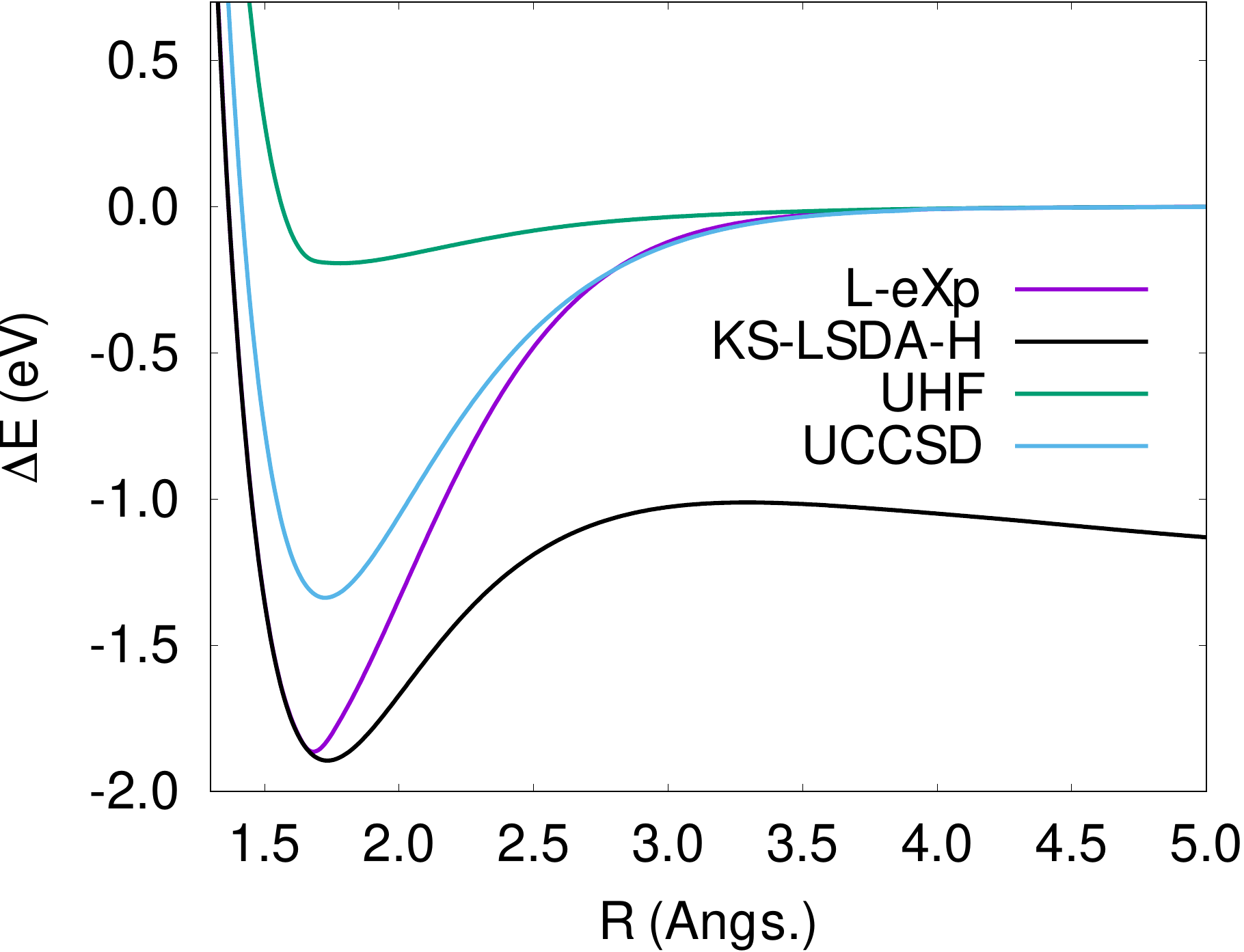}}
\subfigure[Charge of Ne atoms]{\includegraphics[scale=0.4]{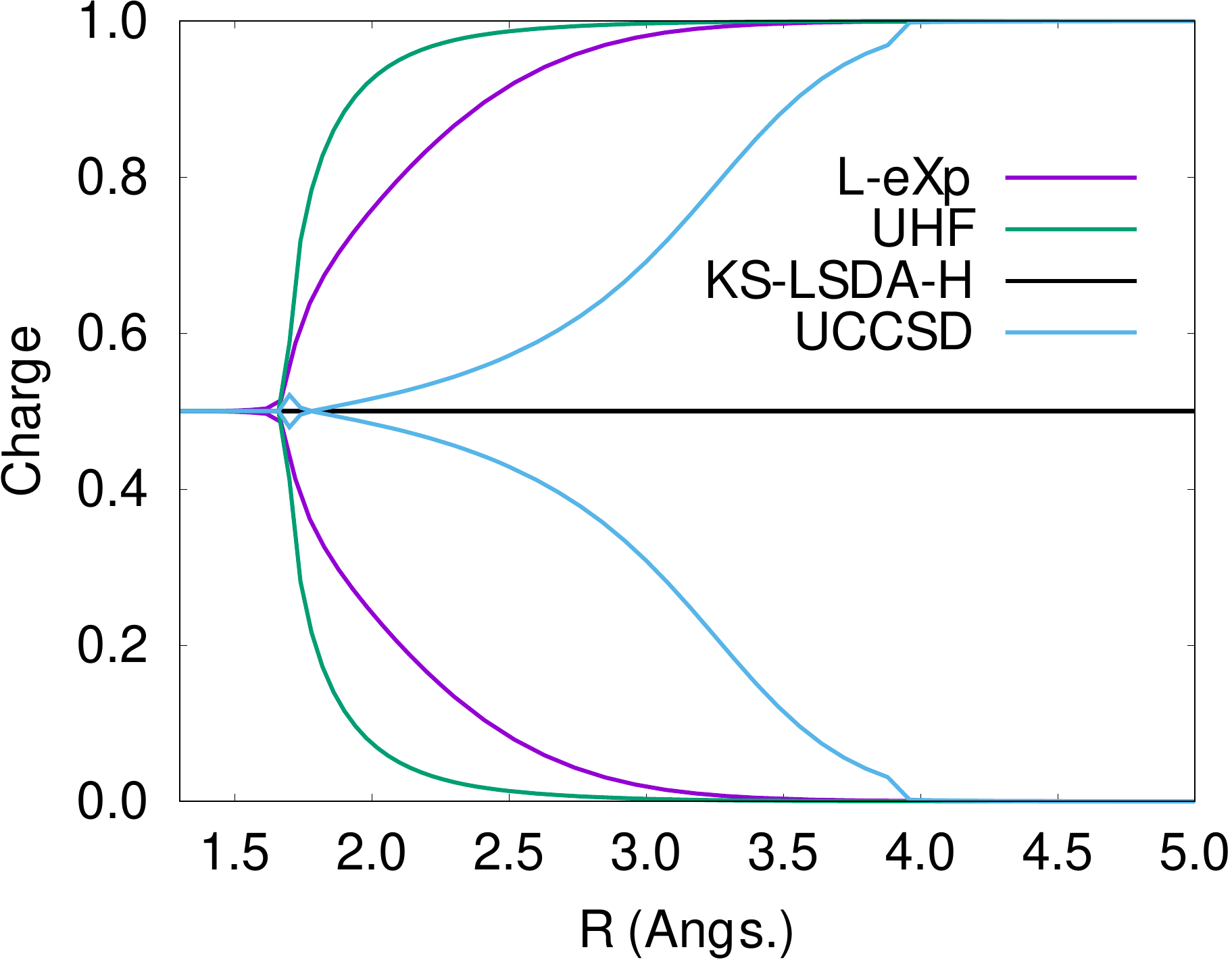}}
\end{center}
\caption{Binding energy of Ne$_2^+$ and its atomic charges computed with different methods: a), Energies determined by 
linearized eXp method (LSDA-H functional), KS LSDA-H, UCCSD, and UHF. b), Atomic charges of each atom as determined by each method 
used in a).}
\label{ne2p_charges}
\end{figure}

In our spin-symmetry breaking approach, at dissociation the left and right atomic systems are
decoupled, so even if the energetics are unfavorable for the neutral configuration, a cluster
amplitude where charge transfer takes place is not possible. For this reason, the charge
delocalization is eliminated in the ground-state calculation. An example of such scenario is the
functional we refer to as ``LSDA-H'' (50 \% HF exchange, 50 \% LSDA exchange, and 100 \% correlation).  Figure
\ref{ne2p_charges}.a shows that the standard KS-LSDA-H technique yields a binding energy that is
quite low at dissociation, due to the fractional-spin errors in the LSDA-H functional. The
self-consistent linearized eXp method corrects this binding energy curve and ensures that the
binding energy meets physical expectation, where it must tend to zero, as in the UHF and UCCSD
results. As opposed to the LSDA-75 functional, LSDA-H in combination with L-eXp overestimates the
binding energy around the equilibrium distance, hence a HF exchange weight between 50 and 75 \%
could give a better result for this matter, or a self-interaction corrected functional such as a
Perdew-Zunger GGA. Further evidence of recovering size-consistency is provided in Figure
\ref{ne2p_charges}.b, where the charge of UHF, L-eXp, and UCCSD tend to the expected symmetry broken
result: one neutral Ne atom, and one Ne cation.  The KS-LSDA-H result is unable to break the spin
symmetry, resulting in the underestimation of the binding energy at long interatomic distances. This
molecular system, Ne$_2^+$ is challenging because it displays both charge and spin entanglement,
quantum effects not encoded by conventional density functional approximations (for other difficult
systems, see Ref. \cite{gould2022poisoning}). Because of this, the conventional approximations, even
though give the right charges, predict erroneous energies and densities (although the charges are
correct). Spin-symmetry is consistent with a collapse of the wavefunction at long distances, hence
better energetics, and can serve as a starting basis for a re-symmetrization (not explored in this
work) consistent with charge and spin entanglement.  Despite the mentioned benefit of spin-symmetry
breaking, in a LSDA-H LR TDDFT (linear response time-dependent DFT) procedure, the state of negative
excitation energy associated to the spurious charge-transfer excitation would return. This is
because of the inherent existence of such state which would manifest in the LR-TDDFT eigenvalue
problem. But for higher amounts of HF exchange this effect can be eliminated, as discussed next for
the LiF system.

We now discuss to the dissociation curve of lithium fluoride, which despite its relative simplicity
as a diatomic molecule, it has been an important system in theoretical chemistry development; there
are fluoride systems that are challenging in DFT method development \cite{gould2022poisoning}. The
unregularized eXp-based self-consistent method can be unstable at moderately long distances between
atoms, not a full dissociation. To understand an underlying reason for this behavior, separate
(non-hybrid) LSDA calculations of the fluorine and lithium atoms show that the lowest unoccupied
orbital of the F atom, with energy -10.3 eV, lies energetically below the highest occupied spin
orbital of the Li atom, which has an orbital energy of -3.2 eV. Therefore, in case the electronic
interaction is weak, the cluster operator during the SCF steps will attempt to transfer an electron
from lithium to fluorine (similarly as in the Ne$_2^+$ case), as it is favorable for the sake of
minimizing the energy. This in turn causes an eventual divergence because the system tries to force
itself to be mostly dominated by a charge-transfer state. Such charge-transfer state of low energy
is eliminated by setting the amount of HF exchange as 75 \%, and by employing the linear or
quadratic eXp method. However, at some intermediate distances between equilibrium and dissociation
it requires $\alpha=0.2$ due to a few energy level differences ($f_{aa}-f_{ii}$) being too close to
zero. We employed a similar value for the hydrogen fluoride in past work, where we showed, again,
that the results remain physically consistent. Even though not tested in this work because of its
unavailability, in our opinion a very appealing improvement in this direction would be the inclusion
of purely density-based self-interaction corrections. 

\begin{figure}[thb]
\begin{center}
\subfigure[]{\includegraphics[scale=0.45]{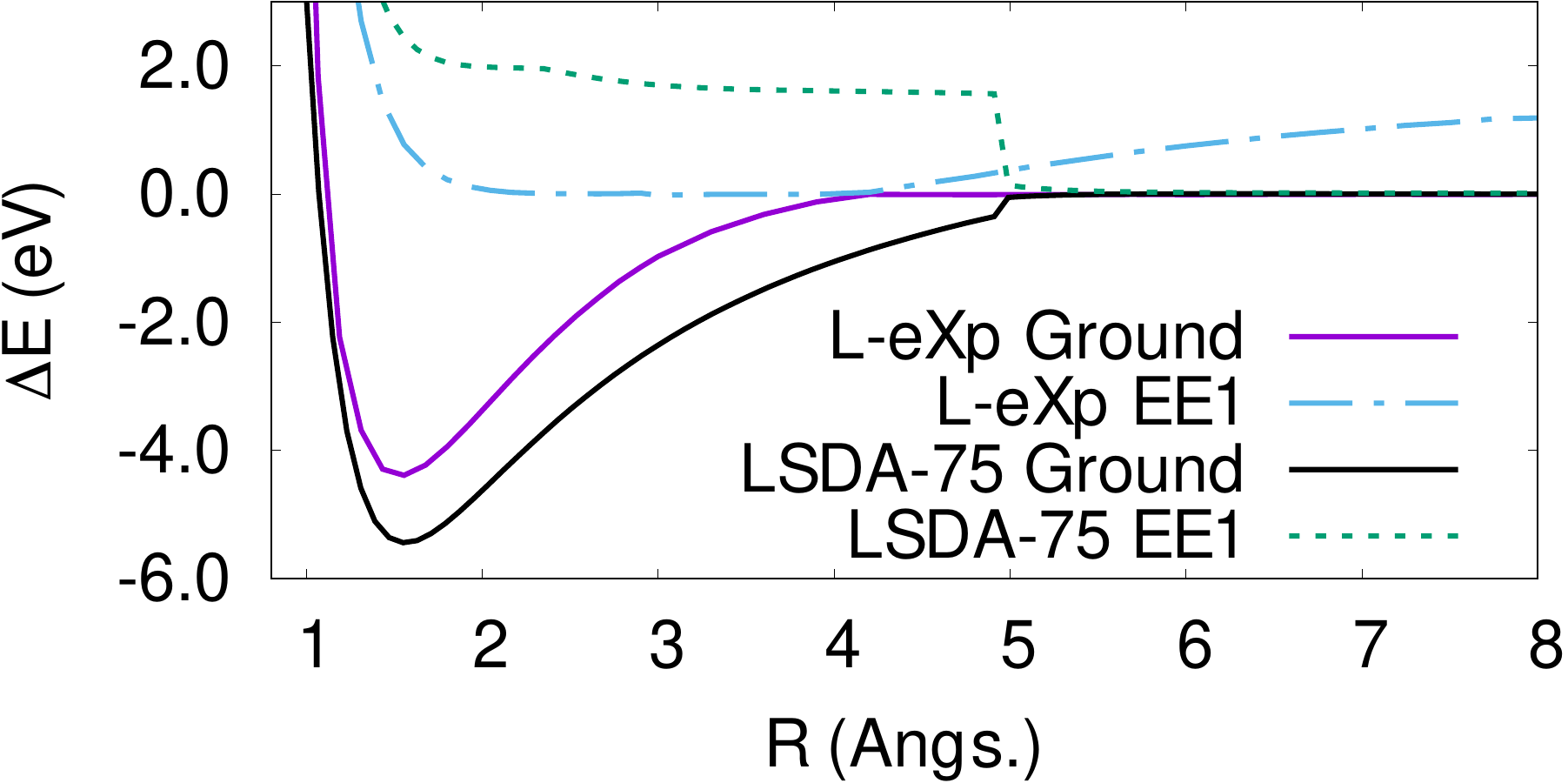}}
\hspace{10pt}
\subfigure[]{\includegraphics[scale=0.55]{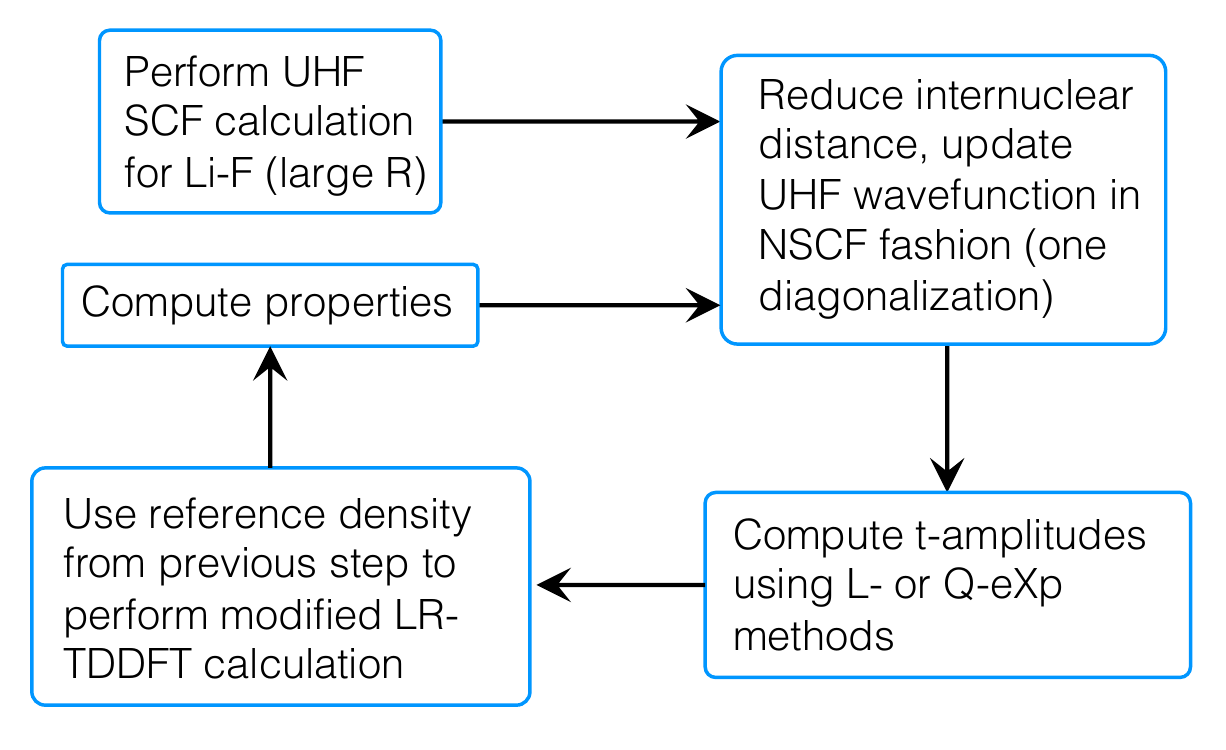}}
\end{center}
\caption{a), Binding energy curves of lithium fluoride computed with the linearized NSCF eXp method vs unrestricted standard KS LSDA-75, including the first excited state for both methods; ``EE1'' stands for the first excited state. a) Steps followed to compute ground- and excited-state energies.}
\label{lif_dissociation}
\end{figure}

To obtain a dissociation curve for LiF our method relies on the UHF reference. At quite long
distances, as expected, the wavefunctions localized correctly. However, examination of the fully
converged UHF solution of LiF reveals a sudden jump in the value of the $\langle S^2\rangle$
operator. The ground-state spin-square $S^2$ value is thus non-analytic at a single point. In
standard hybrid KS-LDA calculations this also introduces a non-differentiable point (or non-unique
force value) in the dissociation curve, as shown in Figure \ref{lif_dissociation}.a. In contrast,
however, in an eXp simulation such jump causes a similar unphysical step in the binding energy curve
because the spin-decomposed exchange energy is sensitive to sudden changes in the spin-densities,
and the solution to the DFT problem is pursued in a post-HF fashion; in other words, if the
spin-density suddenly changes so could the exchange energies. This issue may be resolved if $S^2$ is
either forced to change smoothly, or is maintain fixed. To achieve this, we perform a fully
converged UHF calculation at a relatively long distance, 8 \AA, for example.  The potential energy
curve is then scanned by reducing step-by-step the internuclear distance $R$, where the UHF is
updated only one time, Figure \ref{lif_dissociation}.b.  Such procedure enables us to keep the $S^2$
value of the system  nearly constant. This gives a symmetry-broken reference wave function in which
each atom remains nearly neutral.  Around the equilibrium distance of the system such reference
wavefunction does not correspond to the HF wavefunction of the system. We therefore let the
algorithm update the $t$ amplitudes, but in the linear-response TDDFT step, we include a
contribution from the symmetry-broken reference of the system; such change only requires a minor
modification to the algorithms. Hence, the LR TDDFT solutions produce both the ground and excited
states of the system. In our algorithm then, the LR TDDFT eigenvalues are given with respect to the
reference. An auxiliary wave function in our methodology is of the form
$|\Psi_I\rangle=(X_0^I+\sum_{ai}X_{ai}^I\hat{a}^{\dagger}\hat{i})e^{\hat{t}}|\Psi_0\rangle$, where
$\Psi_0$ is in this case a NSCF UHF wavefunction, and $\mb{X}^I=(X_0^I,\{X_{ai}^I\})$ is the
so-called excitation vector corresponding to state $I$, which can either be the ground or an excited
state. The energies $\{\Omega_I\}$ and vectors $\mb{X}^I$ with respect to the reference are found
solving a LR-TDDFT eigenvalue problem of the form $\mb{A}\mb{X}^I=\Omega_I\mb{X}^I$, where $\mb{A}$
is the Jacobian matrix in the excitation basis.

We find that the standard, symmetry broken (unrestricted), KS-LSDA-75 SCF result for LiF is
size-consistent for the ground-state. However, there are two other issues that are present in this
simulation: First, at dissociation, the charge-transfer configuration is not the first excited
state, as expected, but instead a local excitation of the fluorine atom. Second, the unrestricted
SCF LSDA-75 calculation also features a jump in the value of the squared spin operator, to which the
excitation energies are sensitive too. Hence, even if the local fluorine excitation were ignored, a
sudden jump remains. In Figure \ref{lif_dissociation} we also show the NSCF L-eXp result (however,
the other methods, NSCF L-eXp, SCF Q-eXp, or NSCF Q-eXp yield very similar dissociation energies due
to the need for regularization). The L-eXp ground and first-excited state values are qualitatively
correct. There is a point of near-degeneracy between the ground and excited states, and the charge
transfer excitation is dominant at long distances. The value of this excitation also agrees with FCI
calculations, as well as the fact that the gap between the ground and first excited state is quite
small around the anticrossing point. However, the ground-state binding energy at equilibrium is
underestimated, as well as the internuclear distance at the anticrossing point. Nonetheless, these
features could be fixed, we believe, by an improved density functional approximation, specially
suited for this type of physical situation.

\begin{figure}[thb]
\begin{center}
\includegraphics[scale=0.5]{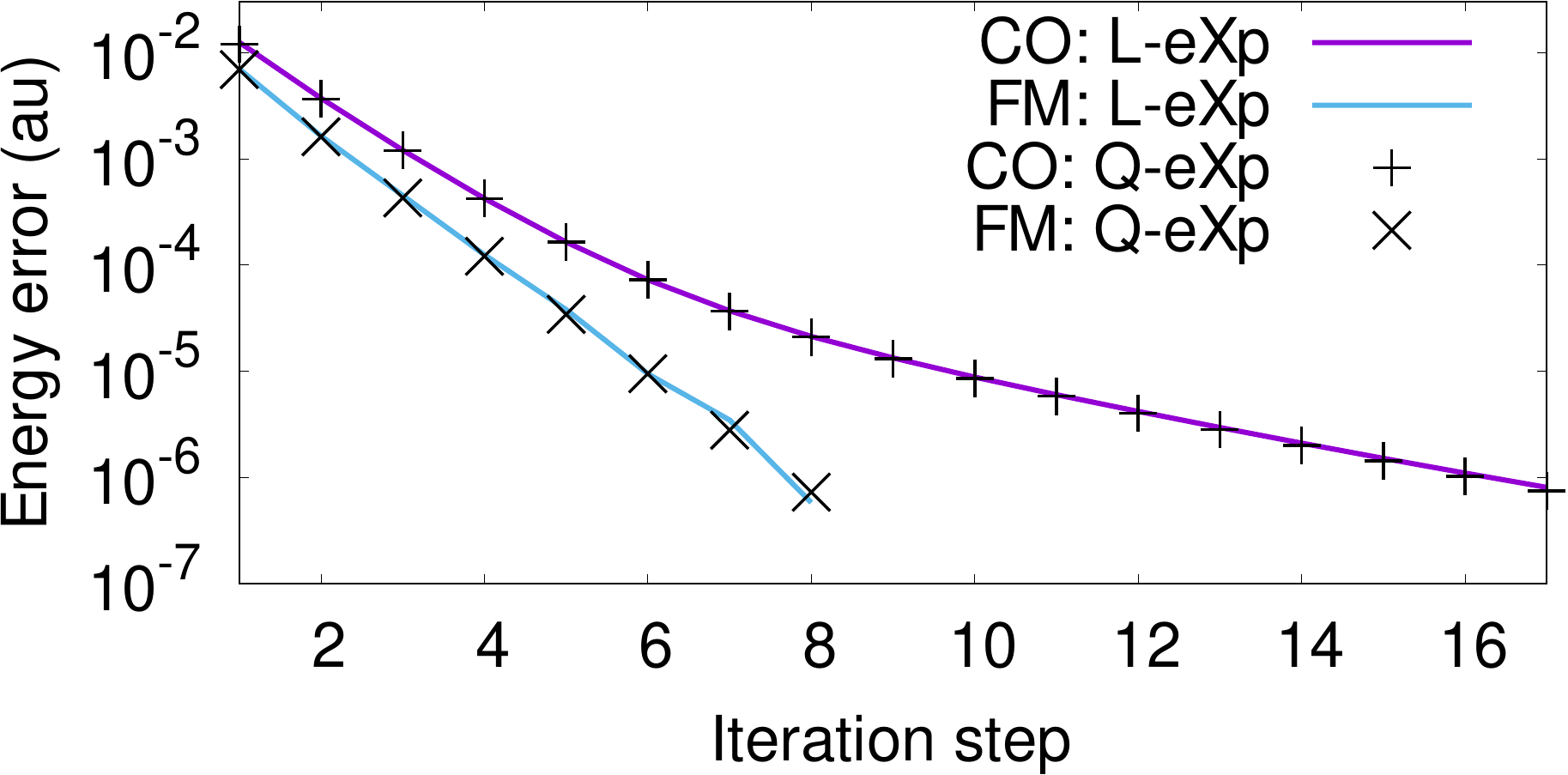}
\end{center}
\caption{Convergence of the linearized and quadratic eXp methods for the the minimum-energy geometry
of carbon monoxide and fluoromethane.}
\label{iter}
\end{figure}

\begin{table*}
  \centering
  \caption{Norm of dipole vectors and absolute value of ground-state energies computed with
  non-regularized L-eXp and Q-eXp, in NSCF and SCF ways, and using the (non-hybrid) LSDA XC
  functional, for a set of molecules at their equilibrium geometries. Values in atomic units.}
  \begin{footnotesize}
    \begin{tabular}{ccc|cc|cc|cc}
          & \multicolumn{2}{c|}{Q-eXp SCF} & \multicolumn{2}{c|}{L-eXp SCF} & \multicolumn{2}{c|}{Q-eXp Non-SCF} & \multicolumn{2}{c}{L-eXp Non-SCF} \\
\cmidrule{2-9}    Molecule & Dipole & Total Energy & Dipole & Total Energy & Dipole & Total Energy & Dipole & Total Energy \\
    \midrule
    $\mr{H_2O}$ & 0.887 & 75.868168 & 0.887 & 75.868168 & 0.882 & 75.868080 & 0.882 & 75.868078 \\
    $\mr{CO}$ & 0.079 & 112.416288 & 0.080 & 112.416288 & 0.496 & 112.398708 & 0.505 & 112.39811 \\
    $\mr{CH_3OH}$ & 0.750 & 114.787583 & 0.749 & 114.787583 & 0.709 & 114.783330 & 0.708 & 114.783279 \\
    $\mr{CH_3F}$ & 0.772 & 138.717566 & 0.772 & 138.717566 & 0.649 & 138.708433 & 0.647 & 138.708284 \\
    $\mr{HCN}$ & 1.190 & 92.610848 & 1.190 & 92.610848 & 1.017 & 92.607709 & 1.014 & 92.607646 \\
    $\mr{H_3O}^{+}$ & 0.677 & 76.137484 & 0.677 & 76.137484 & 0.680 & 76.137398 & 0.680 & 76.137398 \\
    $\mr{OH}^-$ & 0.731 & 75.249223 & 0.731 & 75.249223 & 0.689 & 75.243512 & 0.689 & 75.243331 \\
    $\mr{LiH}$ & 2.211 & 7.911541 & 2.211 & 7.911541 & 2.107 & 7.910796 & 2.105 & 7.910774 \\
    $\mr{LiH_2}^+$ & 1.231 & 8.284455 & 1.230 & 8.284455 & 1.221 & 8.284399 & 1.221 & 8.284399 \\
    \end{tabular}%
\end{footnotesize}
  \label{tab:enerdipoles}%
\end{table*}%

We now discuss the interplay between the different ways to perform calculations: The Q- and L-eXp
methods with and without self-consistency. It is important to remark that even though a calculation
based on cluster operators can be non-self-consistent, the full (self-consistent) Hartree-Fock
orbitals are employed as the starting basis in this present analysis (summarized by Table
\ref{tab:enerdipoles}). As an example, we choose the carbon monoxide and fluoromethane molecules at
the minimum ground-state-energy configuration and consider the convergent behavior of L- and Q-eXp.
Figure \ref{iter} shows that only few steps are required to converge the eXp wavefunctions for an
energy threshold of $10^{-6}$ au. For CO, L- and Q-eXp behave nearly identically, where a quite
small difference is observed for fluoromethane. Our method does not require too many steps because a
Hartree-Fock calculation was performed prior to the SCF eXp simulation. In the molecular set
considered, as shown in Table \ref{tab:enerdipoles}, the self-consistent L- and Q-eXp techniques
perform quite similarly. Some differences are noticeable, however, when the cluster amplitudes are
determined non-self-consistently, especially for CO and CH$_3$F. These two molecules require the
most SCF steps, which correlates well with the differences seen in the NSCF calculations. If
regularization were applied, we would expect fewer SCF steps, but not necessarily more agreement with
the unregularized NSCF calculations, unless these are regularized as well. For this set of molecules
the NSCF step in general improves the properties of interest. This may suggest then that the NSCF
procedures can be of use for practical electronic structure calculations, particularly in cases
where computational savings are needed. An eXp NSCF calculation may also be performed with
regularization, if required. The issue of instabilities in cluster amplitudes is not inherent to
our method only, but it is common in CC theory in general, when energy differences are very small,
for example in semi-metallic systems. But with some form of regularization it can be eliminated.
 
\section{Conclusions}

As an alternative to the standard approach to solve the Kohn-Sham DFT electronic structure problem,
we investigated approximated solutions by means of singles-based cluster operators and amplitudes. These
solutions could serve as a basis for the development of algorithms free of the delocalization error,
with a broader view of electronic excitations, and capable of delivering size consistency in general
(whether the auxiliary Hamiltonian of the system is single- or multi-particle in principle). We found several
potential approaches for the use of this operator in the calculation of alternative single-particle wavefunctions,
thus offering flexible pathways for practical calculations. The linear approach seems to be quite
convenient due its relative accuracy and computational cost. Even though the cluster operator in DFT
can be of use in numerical procedures, its applicability to systems that are inherently of
multireference character (from a WFT point of view) is an unexplored subject, but clearly
encouraging. In this direction, the theoretical procedures presented here may serve
for the formulation of electronic structure models that either couple with automated approaches, or
with techniques that rely on localization/symmetry-breaking of orbitals in order to describe
challenging quantum systems. As it is well-known, different systems need different degrees of
self-interaction corrections. For this reason, connections with quantum embedding
\cite{jones2020embedding} could be beneficial to improve accuracy by assigning different exchange
mixtures to different subsystems in a large molecular system. This, in conjunction with eXp
calculations, could be of interest for density functional calculations with expanded capabilities.
 
\section{Acknowledgment}
G.J. and M.A.M. acknowledge funding from the National Science Foundation Research Experiences for
Undergraduates (REU) Program (award CHE‐1852214). M.A.M and J.M.M.T. thank Montana State University
-- Bozeman for startup support. 

\bibliographystyle{unsrtnat}
\bibliography{refs}
\end{document}